%% file: main.tex
\documentclass[10pt,conference]{IEEEtran}
\input{preamble}

\title{Leveraging Network Methods for Hub-like Microservice Detection}

 \author{\IEEEauthorblockN{Alexander Bakhtin, Matteo Esposito,  Valentina Lenarduzzi, Davide Taibi}

 \IEEEauthorblockA{\textit{University of Oulu}, Finland}

 \IEEEauthorblockA{\{alexander.bakhtin; matteo.esposito; valentina.lenarduzzi; davide.taibi\}@oulu.fi}
 }
\begin{document}

\maketitle

\begin{abstract}
\textit{Context}:
Microservice Architecture is a popular architectural paradigm that facilitates flexibility by decomposing applications into small, independently deployable services. 
Catalogs of architectural anti-patterns have been proposed to highlight the negative aspects of flawed microservice design. 
In particular, the \textit{Hub-like} anti-pattern lacks an unambiguous definition and detection method.

\textit{Aim}:
In this work,  we aim to find a robust detection approach for the Hub-like microservice anti-pattern that outputs a reasonable number of Hub-like candidates with high precision.

\textit{Method}: 
We leveraged a dataset of 25 microservice networks and several network hub detection techniques to identify the Hub-like anti-pattern, namely scale-free property, centrality metrics and clustering coefficient, minimum description length principle, and the approach behind the Arcan tool.

\textit{Results and Conclusion}: 
Our findings revealed that the studied architectural networks are not scale-free, that most considered hub detection approaches do not agree on the detected hubs, and that the method by Kirkley leveraging the Erdős–Rényi encoding is the most accurate one in terms of the number of detected hubs and the detection precision.
Investigating further the applicability of these methods to detecting Hub-like components in microservice-based and other systems opens up new research directions.
Moreover,  our results provide an evaluation of the approach utilized by the widely used Arcan tool and highlight the potential to update the tool to use the normalized degree centrality of a component in the network, or for the approach based on ER encoding to be adopted instead.
\end{abstract}


\section{Introduction}
\label{sec:intro}
\input{Sections/Intro}

\section{Background}
\label{sec:back}
\input{Sections/Background}

\section{Related Work}
\label{sec:RW}
\input{Sections/RelatedWork}

\section{Empirical Study Design}
\label{sec:method}
\input{Sections/Method}

\section{Results}
\label{sec:results}
\input{Sections/Results}

\section{Discussion}
\label{sec:discussion}
\input{Sections/Discussion}

\section{Threats to Validity}
\label{sec:threats}
\input{Sections/Threats}

\section{Conclusion}
\label{sec:conclusion}
\input{Sections/Conclusion}

\bibliographystyle{IEEEtran}
\bibliography{bibliography}
\end{document}

%% file: preamble.tex
\usepackage{graphicx} 
\usepackage{cite}
\usepackage{flushend}
\usepackage{todonotes}
\usepackage{url}
\usepackage[edges]{forest}
\usepackage{float}

\usepackage[many]{tcolorbox}
\definecolor{main}{HTML}{CFCFCF}  
\definecolor{sub}{HTML}{CFCFCF}   

\tcbset{
    sharp corners,
    colback = white,
    before skip = 0.5cm,    
    after skip = 0.7cm      
}

\newtcolorbox{boxCNoTitle}{
    top=10pt,
    rounded corners,
    coltitle=black,
    colframe=gray,
    colbacktitle=sub,
    colback = sub,
    enhanced,
    center,
    boxrule=0.5pt
}
\newtcolorbox{boxC}[2][]{aibox,title=#2,#1}
\tcbset{
  aibox/.style={
    top=10pt,
    boxrule=0.5pt,
    rounded corners,
    coltitle=black,
    colframe=gray,
    colbacktitle=sub,
    colback = sub,
    enhanced,
    center,
    attach boxed title to top left={yshift=-0.1in,xshift=0.15in},
    boxed title style={boxrule=0.5pt,colframe=gray,},
  }
}

\newcounter{keyTakeAwaysCounter}
\newcounter{keyRQAnswerCounter}
\newcounter{keyLimitationsCounter}

%% file: Sections/Intro.tex


Microservice Architecture (MSA) is a popular software development paradigm for scalable and manageable systems \cite{lewis14_microservices}, which facilitates flexibility by decomposing applications into small, independently deployable services. However, with this architecture comes its challenges that can result in design flaws that negatively impact system quality. Catalogs of architectural anti-patterns have been proposed \cite{bakhtin2022survey,cerny2023catalog} to highlight the negative aspects of flawed microservice design \cite{pigazzini2020towards}. Detection of these anti-patterns is crucial in ensuring the quality of microservice-based systems \cite{al2022using}.

Most anti-pattern detection methods employ computationally expensive static and dynamic analyses and, in many cases, do not consider the inter-service dependencies within the MSA. In contrast, modelling the MSA as a network, usually referred to as the Service Dependency Graph (SDG), and applying suitable network methods appears to be an appropriate means of discovering structural outliers within MSA \cite{al2022using, bakhtin2022microservice, bakhtin2025network}. In particular, the centrality metrics can identify those services that occupy central locations in the network and hopefully unveil hidden anti-patterns \cite{bakhtin2025network}.

In this work, we focus on the \textit{Hub-like} anti-pattern, which lacks an unambiguous definition \cite{cerny2023catalog, kirkley2024hubs} and detection approach \cite{azadi2019architectural}, and consider several network hub detection techniques to discover this anti-pattern: scale-free property \cite{barabasi2003scale, barabasi2009scale}, centrality metrics and clustering coefficient \cite{bakhtin2025network}, minimum description length principle \cite{kirkley2024hubs}, and the Fontana et al. \cite{fontana2015automatic} approach leveraged in Arcan \cite{fontana2017arcan}. The goal of this work is to evaluate the agreement of these methods through the Jaccard coefficient and precision through manual validation to guide researchers and practitioners to the most appropriate method for detecting Hub-like microservices.



We leveraged the dataset of 25 SDGs from Bakhtin et al. \cite{bakhtin2025network}.
We discovered that the studied networks are not scale-free, that most considered hub detection approaches do not agree on the detected hubs, and that the ER encoding method by Kirkley \cite{kirkley2024hubs} is the most accurate one in terms of the number of detected hubs and the detection precision.

The contributions of this work are:
\begin{itemize}
    \item A new perspective on Hub-like microservice detection leveraging network science methods;
    \item Dataset of hubs detected in 25 MSA networks;
    \item Comparison of agreement and precision of the considered methods;
    \item Manual validation of Hub-like anti-pattern in 25 MSA networks;
    \item Recommendations on the adoption of hub detection methods for practitioners and researchers.
\end{itemize}



\textbf{Paper Structure.} Section \ref{sec:back} presents the background and the necessary terms for this work while Section \ref{sec:RW} discussed the related work; Section \ref{sec:method} elaborates on the study design and Section \ref{sec:results} presents the results of the study, while Section \ref{sec:discussion} discusses the results and Section \ref{sec:threats} acknowledges the threats to validity of this work, with Section \ref{sec:conclusion} concluding the paper.

%% file: Sections/Background.tex
This section presents the background on Microservice Anti-pattern Taxonomy and network science metrics.

\subsection{Microservice Anti-patterns Taxonomy}
Cerny et al. conducted a tertiary study of MS anti-patterns and provided a \textbf{comprehensive catalog} \cite{cerny2023catalog}, consolidating evidence from seven secondary sources, analyzing 340 primary studies while building upon the earlier work of Palma et al.~\cite{palma2015study} to offer a more up-to-date classification. This taxonomy is grounded in architectural and operational principles specific to microservices and provides \textbf{five} high-level categories:
Intra-service design (9 anti-patterns), Inter-service decomposition (14 anti-patterns), Service interaction (9 anti-patterns), Security (10 anti-patterns), and Team organization (16 anti-patterns).

In particular, Cerny et al. \cite{cerny2023catalog} defined the \textbf{Inter-service decomposition} category as encompassing anti-patterns that emerge because of how services are interconnected, e.g., problematic \textit{integration}, decomposition strategies that hinder \textit{modularity}, and undesirable \textit{service relationships}.
Within this category, we focused on  \textbf{topology} as a subcategory of inter-service decomposition. This subcategory represents the only case where anti-patterns may manifest as measurable structural properties within SDGs. In contrast, the other subcategories require contextual or semantic analysis beyond what network measures can capture. More specifically,
it includes three anti-patterns: Service chain, Cyclic dependency, and Hub-like dependency.

The Service chain and Cyclic dependency anti-patterns have already been studied in many works \cite{walker2020automated, walker2021automated, bushong2021microservice, das2022technical, cerny2022microvision, huizinga2023detecting} and have straightforward detection approaches: \textbf{Service chains} in the SDG can be listed via a Breadth-First search algorithm and sorted by length to detect the longest ones \cite{bakhtin2022microservice, bushong2021microservice, das2022technical, cerny2022microvision, cerny2023catalog}. Furthermore, \textbf{Cycles} can be detected with the Depth-First search algorithm, and this anti-pattern is considered in most works dealing with architectural anti-patterns \cite{walker2020automated, walker2021automated, bushong2021microservice, das2022technical, al2022using, gortney2022visualizing, huizinga2023detecting, cerny2024static}. However, the notion of a Hub-like service is more subjective, and few works have attempted to detect this anti-pattern, with most relying on the same detection method (see below). For these reasons, we aim to explore different network science techniques applicable to detect Hub-like microservices in this work.

\textbf{Hub-like dependency.} Cerny et al. \cite{cerny2023catalog} reference Azadi et al. \cite{azadi2019architectural} to define the anti-pattern as: \emph{``A service has (outgoing and ingoing) dependencies with a large number of other services. The service becomes a central point of dependency for many other services".} While Cerny et al. \cite{cerny2023catalog} only name the anti-pattern as \emph{Hub-like dependency} without alternative namings, Azadi et al. \cite{azadi2019architectural} actually noted additional names encountered in previous literature, namely \textit{Hub-like modularization} \cite{sharma2016designite} and \textit{Link Overload} \cite{le2016relating}.

Fontana et al. \cite{fontana2017arcan} consider the total degree of the class usage graph as the detector for Hub-like anti-pattern and provide the Arcan tool capable of detecting this anti-pattern.
Capilla et al. \cite{capilla2023detecting}, Bacchiega et al. \cite{bacchiega2024refactoring}, Pigazzini et al. \cite{pigazzini2022exploiting} leveraged the Arcan tool \cite{fontana2017arcan} for detecting the Hub-like microservices.
Lino et al. \cite{lino2024musvision} provide a tool for Microservice Anti-pattern detection, including Hub-like services detection. They define the Hub-like service as having \emph{too many} connections to others. In practice, they identified two services as Hub-like for having 4 connections to other services.
The Designite tool by Sharma et al. \cite{sharma2016designite} requires both the fan-in and fan-out, i.e., in-degree and out-degree, of a component to be above 20 \cite{azadi2019architectural}.
ARCADE \cite{le2016relating} requires that the total amount of dependencies, i.e., the degree, is higher than the mean plus standard deviation across the entire system \cite{azadi2019architectural}.
Silva et al. \cite{silva2024mfe} consider the Hub-like dependency in the context of Micro-Frontend anti-patterns.
In her industrial conference talk, Watt \cite{watt2019using} proposed to use degree centrality to detect services with too much communication.

\input{Tables/methods}

\subsection{Network science}

In this section, we provide the relevant definitions and concepts from network science for our work: degree, centrality metrics, clustering coefficient, scale-free property, and minimum description length principle.

\textbf{Degree, In-degree, Out-degree.}
In network science, the \textit{degree}  of a specific node (microservice) is the total number of connections that node has. More specifically, the \textit{in-degree} is defined as the total amount of incoming connections, and the \textit{out-degree} as the number of outgoing connections. In the software architecture domain, they are known respectively as the Fan-in and Fan-out of a component.
In MSA literature, in the context of SDGs, they are known as the Absolute Importance of a Service (AIS), and the Absolute Dependence of a Service (ADS) \cite{rud2006product}.
These values thus exist on the absolute scale.
Engel et al. \cite{engel2018evaluation} use network degree to evaluate several MSA principles (patterns).

\textbf{Degree, In-degree, Out-degree centrality.}
Since the plain degree values exist on the absolute scale, it is customary in network science to also define the degree \textit{centrality} as the total amount of connections of a node divided by the total amount of \textit{possible} connections, i.e., all the rest of the nodes. Similarly, \textit{in-degree centrality} is the total amount of incoming connections divided by all possible connections, and \textit{out-degree centrality} divides the total amount of outgoing connections. These centrality metrics are thus confined to the $[0.0; 1.0]$  interval, which makes it easier to analyze the values irrespective of the size of the network (Figure \ref{fig:comparison}).

\textbf{Betweenness and Closeness centrality.}
Some centrality scores are related to the \emph{connectivity} of the network and consider the possible paths within it.
\textit{Betweenness centrality }of a node is defined as the fraction of the shortest paths between all other services that pass through that node and all existing shortest paths. It thus measures how critical a node is in the connectivity and communication in the network. \textit{Closeness centrality} of a node is inversely proportional to the distances of that node to all other nodes, thus measuring the node's reachability from all others.
Gaidels and Kirikova \cite{gaidels2020service} and Farsi \cite{farsi:hal-03825330} have discussed the feasibility of using these centrality metrics to analyze the SDG. However, they did not associate the metrics or their values with anti-patterns, proposing it for future work.
Borges and Khan \cite{borges2019algorithm} describe the network of class imports in a microservice system using Betweenness and Closeness centralities.

\textbf{Eigenvector, PageRank, Hub and Authority centrality.}
Eigenvector centrality is calculated by identifying the principal eigenvector of the network adjacency matrix. Such an approach allows each node to propagate its centrality value to other connected nodes, thus influencing the centrality of other nodes in the network. Several transformations of the adjacency matrix provide related centrality metrics: \textit{PageRank} centrality considers how likely a node is to be encountered by a random walk on the network, and \textit{Hub and Authority} scores are a coupled pair of metrics, initially proposed for citation networks: a node with high hub score should be the one pointing to many nodes with high authority score, and a node with high authority score should be the one referred to by many hubs with high hub score.

\textbf{Clustering coefficient.}
The clustering coefficient measures how grouped the nodes in the network are. For each node, it considers all possible pairs of neighbors and determines the fraction that are connected. A node with a high coefficient is thus the one among a tightly connected group of nodes, and a node with a low coefficient connects mostly isolated nodes.

\textbf{Scale-free property.}
Scale-free networks exhibit a distribution of node degrees that follows a power law. This property was discovered through mapping the World Wide Web network \cite{albert1999diameter}, and networks that possess the scale-free property are known to contain several hub nodes that accumulate many connections \cite{barabasi2003scale}. Networks from many domains have been shown to possess the scale-free property \cite{barabasi2009scale}. To our knowledge, no work has attempted to assess MSA for exhibiting scale-free property.

\textbf{Minimum Description Length Principle.}
Alec Kirkley proposed to detect hubs in networks using the minimum description length principle \cite{kirkley2024hubs}. This method is based on the observation that network information can be compressed and transferred in several ways, requiring different amounts of bits. In particular, simply enumerating all nodes and edges results in a length of encoding that is not optimal. The number of bits needed can be reduced if several nodes accumulating many connections, i.e., hubs, are selected and transferred first. Kirkley \cite{kirkley2024hubs} thus considers two different encodings based on hubs - Erdős–Rényi (ER) and Configuration Model (CM), and finds the optimal selection of hubs by minimizing the resulting amount of bits necessary to transfer the network. According to Kirkley, the advantage of this method over many others is that it is completely non-parametric.
Moreover, he compared the results to two baseline approaches - Avg and Loubar. In the Avg (Average) method, a node is considered a hub if its degree exceeds the average degree $\langle k \rangle = \frac{E}{N}$, where $E$ is the total number of edges and $N$ is the total number of nodes. According to the Loubar method, a node is considered a hub if its degree falls into $1 - \frac{\langle k \rangle}{max(k)}$ quantile, where $max(k)$ is the maximum degree observed in the network.

%% file: Tables/methods.tex
\begin{table*}[t]
\centering
\caption{Current applications of centrality to Hub-like anti-pattern detection in literature\\(PageRank, Hub and Authority score not used or mentioned anywhere)}
\label{tab:methods}
\resizebox{\linewidth}{!}{%
\begin{tabular}{l|l|l|l|l|l|l|l|l|l} \hline 
                & \textbf{Degree} & \textbf{In-degree} & \textbf{Out-degree} & \textbf{Degree c.} & \textbf{In-degree c.} & \textbf{Out-degree c.}& \textbf{Betweenness} & \textbf{Closeness} & \textbf{Eigenvector} \\ \hline
Leveraged &\cite{fontana2017arcan, watt2019using, le2016relating}&\cite{lino2024musvision, sharma2016designite}&\cite{lino2024musvision, sharma2016designite}&-&-&-&-&-&-\\
Only mentioned &\cite{engel2018evaluation,gaidels2020service,farsi:hal-03825330}&\cite{engel2018evaluation,gaidels2020service}&\cite{engel2018evaluation,gaidels2020service}&\cite{bakhtin2025network}&\cite{bakhtin2025network}&\cite{bakhtin2025network}&\cite{gaidels2020service,farsi:hal-03825330,borges2019algorithm,bakhtin2025network}&\cite{gaidels2020service,farsi:hal-03825330,borges2019algorithm,bakhtin2025network}&\cite{gaidels2020service,farsi:hal-03825330,bakhtin2025network} \\ \hline 
\end{tabular}%
}
\end{table*}

%% file: Sections/RelatedWork.tex
\begin{figure*}[h]
    \centering
    \includegraphics[width=0.9\linewidth]{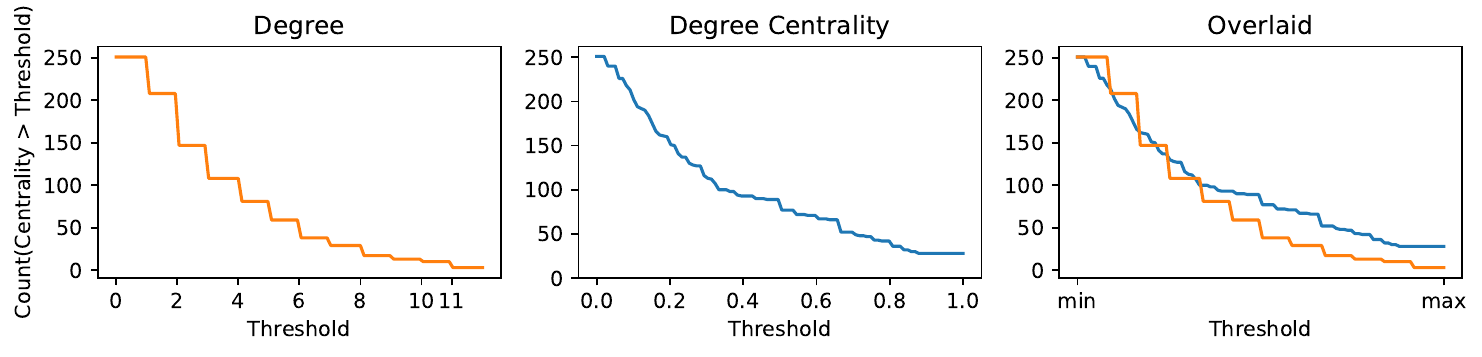}
    \caption{Comparison of Cumulative Distributions of Degree and Degree centrality}
    \label{fig:comparison}
\end{figure*}

In this section, we discuss related works identifying Hub-like microservices and, generally, detecting hubs in networks.

As described in Section \ref{sec:back}, the only works we identified that proposed a way to detect the Hub-like pattern utilize the node degree \cite{fontana2017arcan, lino2024musvision, azadi2019architectural}.
However, this approach suffers from several concerns, which are also highlighted in Watt's talk \cite{watt2019using} and review by Azadi et al. \cite{azadi2019architectural}:

\textbf{Degree is not normalized.}
If the analysis is based on the value of the degree itself, analyzing systems and networks of different scales becomes complicated. Degree centrality, on the other hand, is normalized with respect to network size, and thus makes consistent analysis of distinct systems possible (Section \ref{sec:back}). Figure \ref{fig:comparison} compares the absolute and normalized degree centralities of data collected for this paper to demonstrate the problems using the absolute scale.

\textbf{The choice of threshold is unreliable.}
Most authors seem to identify a certain number of connections as \emph{too many}, which is subjective, and, similarly to the first concern, depends on the scale of the system. Azadi et al. \cite{azadi2019architectural} noted that there is no agreement on the threshold adopted for the detection of Hub-like anti-pattern.
Lavazza and Morasca \cite{lavazza2016identifying, morasca2016slope} argue for careful choice of thresholds for software metrics, noting that frequently, either only knowledge about the distribution of metric or the experience and confidence of the practitioners are used to set the thresholds.
Fontana et al. \cite{fontana2015automatic} proposed a quartile-based approach for selecting the threshold of metrics; however, they are still assessing metrics on the absolute scale, assuming they gathered a representative sample of systems to analyze.
ARCADE \cite{le2016relating} requires that the total amount of dependencies, i.e., the degree, is higher than the mean plus standard deviation across the entire system, which is a strategy usually employed for the bell-shaped distributions, while the degree of components in a dependency network is unlikely to follow such a distribution \cite{kirkley2024hubs, barabasi2003scale}. The Designite tool \cite{sharma2016designite} requires both the Fan-in and Fan-out of a component to be above 20 \cite{azadi2019architectural}.

\textbf{Other aspects of the network are not considered.}
Different aspects of the MSA network can be considered to determine whether the service is Hub-like or not. Section \ref{sec:back} presented several definitions of centrality metrics that take different topological properties into account, like consolidating paths or pointing to influential nodes. Works that do discuss and compute other centralities (\cite{gaidels2020service,farsi:hal-03825330, borges2019algorithm}) do not apply them to detect a particular pattern but rather just consider the service ranking w.r.t the computed centralities, labeling the most central services as potentially problematic (Table \ref{tab:methods}).
Bakhtin et al. \cite{bakhtin2025network, bakhtin2025ccp} recently presented network centrality as a new perspective on microservice architecture and proposed it to be used for anti-pattern detection. Kirkley \cite{kirkley2024hubs} on the other hand, argues against using degree, degree centrality or any other centrality-based ranking, since it requires either setting a threshold or defining the notion of \emph{too high}/\emph{too many} \cite{esposito2023can}, usually through an outlier detection method \cite{falessi2023enhancing}, which comes with its own set of parameters. Instead, he argues for developing completely non-parametric methods for hub detection.

To our knowledge, this is the first attempt to consider the detection of Hub-like anti-pattern in microservice systems from a network science point of view.

%% file: Sections/Method.tex
\begin{figure*}
    \centering
    \includegraphics[width=\linewidth]{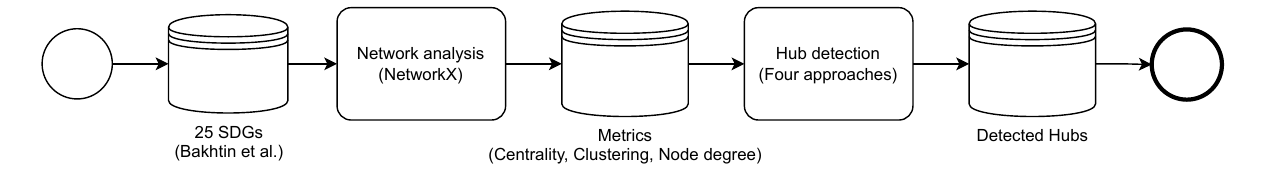}
    \caption{Data collection and analysis process}
    \label{fig:bpmn}
\end{figure*}

In this section, we describe the design of the empirical study following the guidelines by Wohlin et al.~\cite{wohlin_experimentation_2024} (Figure \ref{fig:bpmn}). 

\subsection{Goal, Motivation and Research Questions}


We are interested in seeing if different network methods applicable to the detection of network hubs enable us to detect Hub-like microservices, as defined by Cerny et al. in \cite{cerny2023catalog}.

Therefore, we formulate the following research questions (RQs):

\begin{boxC}{\textbf{RQ$_1$}}
Which network and centrality methods detect the Hub-like microservices?

  \begin{itemize}
      \item \textbf{RQ$_{1.1}$}: Do microservice networks possess the scale-free property, thus containing emerging hubs?
      \item \textbf{RQ$_{1.2}$}: Can methods based on the minimum description length principle identify Hub-like microservices?
      \item \textbf{RQ$_{1.3}$}: Can centrality metrics coupled with clustering coefficient identify Hub-like microservices?
      \item \textbf{RQ$_{1.4}$}: How well does the approach from the Arcan tool detect Hub-like microservices?
  \end{itemize}
\end{boxC}

We are interested in investigating whether MSA networks follow a \textbf{scale-free topology}, where degree distributions follow a power-law. In such graphs, hubs are known to be emergent structural properties, explained, for example, by preferential attachment \cite{newman2001clustering}. Assessing this with statistical fitting can allow us to leverage known properties of scale-free networks for hub detection.

Kirkley \cite{kirkley2024hubs} has discussed current approaches for hub detection, noting that most of them are based on thresholding node properties such as degree. He instead proposed two methods based on \textbf{minimum description length principle}, that is, the observation that network information can be compressed to a smaller amount of bits if certain nodes with many connections are selected, i.e., hubs. The methods then try to label the optimal set of nodes as hubs to minimize the number of bits necessary to encode the network. We are interested in observing whether microservices detected with such a method are exhibiting the Hub-like anti-pattern.

Several works have proposed detecting problematic microservices by ranking them concerning centrality metrics \cite{gaidels2020service, borges2019algorithm}. We can thus leverage \textbf{centrality for the detection of the Hub-like} anti-pattern. However, a node can have a high centrality value even if it is within a densely connected set of nodes. We can utilize the \textbf{clustering coefficient} to filter nodes that bring together other disconnected nodes and thus represent a hub, since the clustering coefficient measures the fractions of neighbors of a given node that are connected, and thus is expected to be low for network hubs.

Previous works that detected the Hub-like anti-pattern \cite{capilla2023detecting,bacchiega2024refactoring,pigazzini2022exploiting} leveraged the Arcan tool \cite{fontana2017arcan}.
Arcan uses a quartile-based approach of setting a threshold on the Fan-in and Fan-out of the class or package to label it as Hub-like \cite{fontana2015automatic}. However, there are several shortcomings in this method: (1) The method relies on the absolute amount of connections, which can be misleading if studied systems are of different scale; (2) The threshold is chosen based on a sample of benchmark systems considered representative; (3) The tool applies the detection to a class or a package and not the overall microservice. We are thus interested in replicating the same threshold calculation approach on the degree of a microservice in the SDG. Moreover, due to considerations discussed in Section \ref{sec:RW}, we aim to see whether the normalized degree centrality provides a more robust way to detect the Hub-like anti-pattern than the absolute degree. We will thus also apply the Arcan approach to the degree centrality metric. We cannot use the Designate method, since it uses an absolute threshold of the Fan-in and Fan-out of a class which cannot be migrated to the inter-service level, while the ARCADE method is very similar to the considered Loubar method, while suffering from the concerns mentioned in Section \ref{sec:RW}.

\begin{boxC}{\textbf{RQ$_2$}}
What is the agreement of detected Hub-like microservices among the scale-free, network-based, centrality-based, and Arcan methods?
\end{boxC}
We are interested in observing whether the different methods agree about which microservices are Hub-like. In case of major disagreement, it would highlight the importance of selecting and adopting the appropriate method to detect this anti-pattern, since otherwise, different researchers and practitioners might end up with a serious gap in understanding each other.

\begin{boxC}{\textbf{RQ$_3$}}
What is the precision of detecting Hub-like microservices of the scale-free, network-based, centrality-based, and Arcan methods?
\end{boxC}
There is no public dataset of microservices labeled as Hub-like. For this reason, we leverage manual validation to determine which services are Hub-like.
Since there is a large number of data points and the notion of Hub-like anti-pattern is subjective, it is unfeasible to evaluate all microservices for the presence of Hub-like anti-pattern and thus calculate also recall.
We approach this issue from the practitioners' point of view and thus wish that a method that performs analysis of the Hub-like microservice in a production system outputs only the services that would be of interest to inspect to the senior architect, and does not provide too many services that would be immediately rejected by the architect. Thus, we are interested in maximizing the Precision of the output. Selecting a method that maximizes the Recall would bias the results towards the present authors' understanding of what to consider a Hub-like anti-pattern, which might be different from the perspective of an architect of a real MSA project.

\subsection{Study Context}

We used the dataset of microservice architectural networks provided by Bakhtin et al. \cite{bakhtin2025network}.
The authors compiled a dataset of OSS MSS projects suitable for network analysis by reconstructing their SDGs. They leveraged \textit{Code2DFD} tool \cite{Code2DFD23, schneider2024comparison} to reconstruct the architecture statically, which only applies to the Java Spring framework. They thus identified 125 Java Spring systems from four datasets of OSS MSS \cite{imranur2019curated, schneider2023microsecend, amoroso2024dataset, yang2024feature}. These 125 Java Spring projects provide complete architectures of 24 OSS MSS projects reconstructed successfully with the \textit{Code2DFD} tool.
We took the 24 SDGs directly from \cite{bakhtin2025network}.
Moreover, we took an additional version of the \emph{train-ticket} benchmark project to enrich our data from \cite{bakhtin2025ccp}
since the architecture varies noticeably between releases \textit{v0.0.1} and \textit{v1.0.0} of the project, and thus can provide additional data points for analysis.
The list of the analyzed projects in available in the Replication Package \cite{replication}.

\subsection{Data Collection}
We collected the following network metrics and coefficients with the NetworkX Python package: 
\begin{itemize}
\item \textbf{Centrality metrics}   (Degree centrality; In-degree centrality; Out-degree centrality; Betweenness centrality; Closeness centrality; Eigenvector centrality; PageRank centrality; Hubs and Authority scores)
\item \textbf{Clustering coefficient}
\item \textbf{Node degrees} (Degree; In-degree; Out-degree)
\end{itemize}

We selected the aforementioned centrality metrics and the clustering coefficient for RQ$_{1.3}$ because they provide the values normalized to the $[0.0;1.0]$ interval, thus having the potential to provide a universal method for detecting Hub-like microservices decoupled from the scale and size of the studied system; node degrees are required for RQ$_{1.1}$ and RQ$_{1.4}$, while RQ$_{1.2}$ only requires the network as input.

\subsection{Data Analysis}
We have computed nine centrality metrics, node degrees, and the clustering coefficient for 251 microservices from 25 SDGs (Figure \ref{fig:bpmn}).

\subsubsection{Scale-free property (RQ$_{1.1}$)}
We tested the Scale-free property of the studied SDGs by testing the goodness of fitting a power law to the distribution of the microservices' degree, out-degree, and in-degree, as commonly done in network science \cite{barabasi2009scale}. We leveraged the Python package \texttt{powerlaw}\footnote{\url{https://pypi.org/project/powerlaw/}} for this analysis. We chose $0.01$ as the level of significance, as was done by Bakhtin et al. \cite{bakhtin2025network} when analyzing this dataset.







\subsubsection{Minimum Description Length Principle (RQ$_{1.2}$)}
Kirkley \cite{kirkley2024hubs} provided the code\footnote{\url{https://github.com/aleckirkley/Network-hubs}} implementing the two methods based on the minimum description length principle as well as the baseline Average Degree and Loubar approaches. We leverage this code for our analysis.
Given the input network, these algorithms list nodes detected as hubs. The assignment is boolean, i.e., there is no "strength" of being a hub given to the nodes. The algorithms base their analysis of a particular node on either the incoming or outgoing connections. Thus, each algorithm can provide two different lists of hubs.

\subsubsection{Clustering coefficient and Centrality (RQ$_{1.3}$)}
We aim to combine a centrality metric and the clustering coefficient and consider them jointly to detect Hub-like services since a node with a high centrality can exist in a highly connected, dense network, so it would not be a candidate for a Hub-like service. Such a case can be distinguished with the clustering coefficient - a service with a high clustering coefficient would be in the middle of the densely connected region \cite{zhou2005maximal}, so we should ignore it. Conversely, a proper hub in a network where connections emerge radially from the hub would have a high centrality score but a low clustering coefficient since most of the neighbors of such a hub do not connect \cite{zhou2005maximal}.

To determine such cases, we plot the centrality score vs. clustering coefficient on a scatter plot (e.g., Figure \ref{fig:closeness}) and find the points corresponding to services in the top-left corner, indicating high centrality and low clustering coefficient.
Such an approach can evaluate the strength of a specific microservice being considered a hub, indicating the position on the top-left to bottom-right corner diagonal of the scatter plot. The strength can be computed as $centrality - clustering$, i.e., the difference of the two metrics, provided both are constrained to $[0.0,1.0]$ interval (Figure \ref{fig:closeness}, diagonal lines). This strength measure ranges thus from $-1.0$ to $1.0$.
Two authors manually analyzed the scatter plots, selected candidate points, and reached consensus in case of disagreements.

 We analyzed the agreement via Cohen's $\kappa$ \cite{cohen_coefficient_1960}. 
The kappa statistic is commonly used to evaluate the agreement between raters or classifiers (IRA). We can use Cohen's $\kappa$ to measure the level of agreement between two independent authors who categorized the datasets according to different characteristics. $\kappa$ provides a metric for the reliability and consistency of the categorizations (Table \ref{tab:kappa-agreement}).

\input{Tables/kappa}

\subsubsection{Arcan Approach (RQ$_{1.4}$)}

Fontana et al. \cite{fontana2015automatic} proposed a quartile-based approach for automatic selection of thresholds for software metrics, which they integrated into the Arcan tool \cite{fontana2017arcan} for detecting the Hub-like anti-pattern based on the combined amount of incoming and outgoing connections, i.e the degree of the component. We replicate the same approach to determine the threshold of the degree of a microservice to detect it as Hub-like. Moreover, we also use degree centrality as a different metric for choosing the threshold, according to arguments presented in Section \ref{sec:RW}, to compare the two approaches.

\subsubsection{Comparison of methods (RQ$_{2}$)}

We split the hub detection methods into three groups based on the kind of connections they consider - incoming, outgoing, or both (Table \ref{tab:groups}).
We can compute the pairwise agreement of the methods with the Jaccard coefficient in each of the groups, as well as the agreement among each entire group with the Fleiss Kappa coefficient \cite{falotico2015fleiss}, which measures agreement between several \emph{raters}, i.e., hub detectors, assigning microservices to different \emph{categories}, i.e., Hub-like and non-Hub-like. 

\input{Tables/groups}
\subsubsection{Manual validation (RQ$_{3}$)}

Two authors inspected the microservices detected as hubs with each method and the corresponding SDG and judged whether the service represents an instance of the Hub-like anti-pattern. We also considered the role of the service and the services it is connected to, as determined by the names and possibly documentation.
We defined and labeled separately the \emph{infrastructural} hubs (IH), i.e., instances of microservices such as \textit{eureka}, \textit{consul}, \textit{configuration server}, etc.
Such services are required to connect to every other service due to their functionality and, thus, do not require the attention of the lead architect. However, most network-based methods would still detect them as hubs due to the many connections such services have, so we should still consider them as hubs for manual validation. We can calculate the Precision in three ways: treating such infrastructural hubs as True Positives, ignoring such services, and treating IH as False Positives. This allows us to see whether methods can naturally filter infrastructural hubs from the output.

%% file: Tables/kappa.tex
\begin{table}[tb]
\centering
\caption{Interpretation categories for agreement levels by $\kappa$ value}
\label{tab:kappa-agreement}
\resizebox{0.95\linewidth}{!}{%
\begin{tabular}{c|c|c|c|c}
\hline
$\kappa < 0$ & $0 \leq \kappa < 0.4$ & $0.4 \leq \kappa < 0.6$ & $0.6 \leq \kappa < 0.8$ & $0.8 \leq \kappa < 1$ \\
\hline
None & Poor  & Discrete  & Good  & Excellent  \\
\hline
\end{tabular}%
}
\end{table}

%% file: Tables/groups.tex
\begin{table}[]
\caption{Categories of hub detection approaches based on considered connections}
\scriptsize
\label{tab:groups}
\centering
\begin{tabular}{l|l|l}
\hline
Incoming Connections      & Outgoing Connections       & All Connections            \\ \hline
Avg\_in         & Avg\_out          & Degree centrality       \\
Loubar\_in         & Loubar\_out         & Arcan\_abs            \\
CM\_in               & CM\_out               & Arcan\_norm         \\
ER\_in              & ER\_out           & Betweenness centrality  \\
In-degree centrality  & Out-degree centrality  & Closeness centrality    \\
Authority score     & Hub score            & PageRank centrality  \\
 Eigenvector centrality  &                           &  \\ \hline
\end{tabular}
\end{table}

%% file: Sections/Results.tex
In this section, we described the results of applying each method for hub detection (RQ$_{1.1}$ - RQ$_{1.4}$), the agreement between the methods (RQ$_{2}$), and precision concerning the ground truth (RQ$_{3}$). The list of services labeled as hubs by each method is available in the Replication Package \cite{replication}.

\subsection{Scale-free property (RQ$_{1.1}$)}

\begin{figure*}
    \centering
    \includegraphics[width=0.9\linewidth]{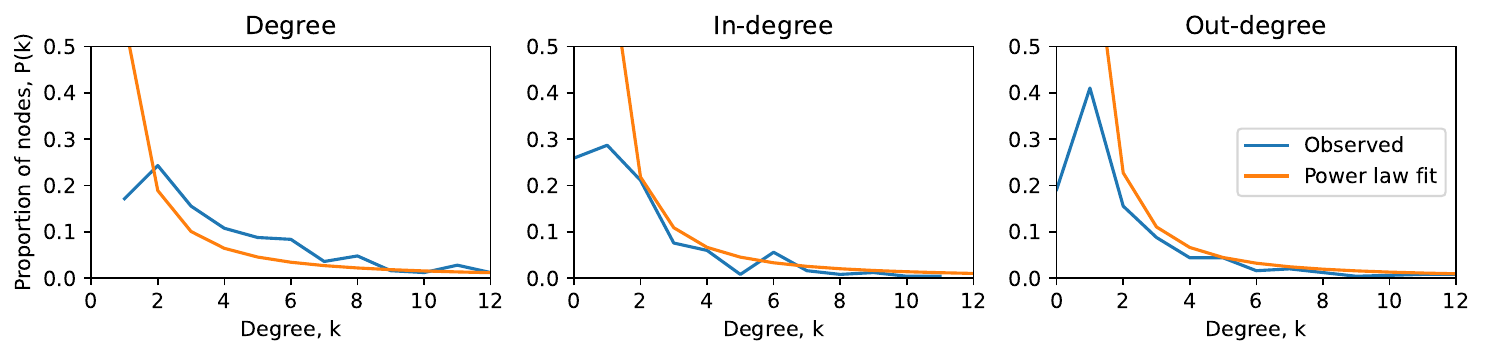}
    \caption{Power law fits of the distributions of node degrees to determine the scale-free property}
    \label{fig:scalefree}
\end{figure*}

We have assessed the distributions of degree, out-degree, and in-degree of microservices for the scale-free property, i.e., exhibiting a power-law (Figure \ref{fig:scalefree}).
The p-values of the tests for degree, in-degree, and out-degree were $0.035$, $0.053$, and $0.055$, respectively.
Since for all three cases, the p-value was above $0.01$, we conclude that \textbf{studied microservice systems are not scale-free networks}, so we cannot use properties of scale-free networks to reason about the hubs.

\subsection{Minimum Description Length Principle (RQ$_{1.2}$)}

We performed hub detection using all four methods discussed by Kirkley~\cite{kirkley2024hubs} both for incoming and outgoing connections, thus obtaining \textbf{eight different lists of hub candidates}: Average\_In, Average\_Out, Loubar\_In, Loubar\_Out, CM\_in, CM\_out, ER\_in, ER\_out (Table \ref{tab:groups}).

\subsection{Clustering Coefficient (RQ$_{1.3}$)}

\begin{figure}
    \centering
    \includegraphics[width=0.9\linewidth]{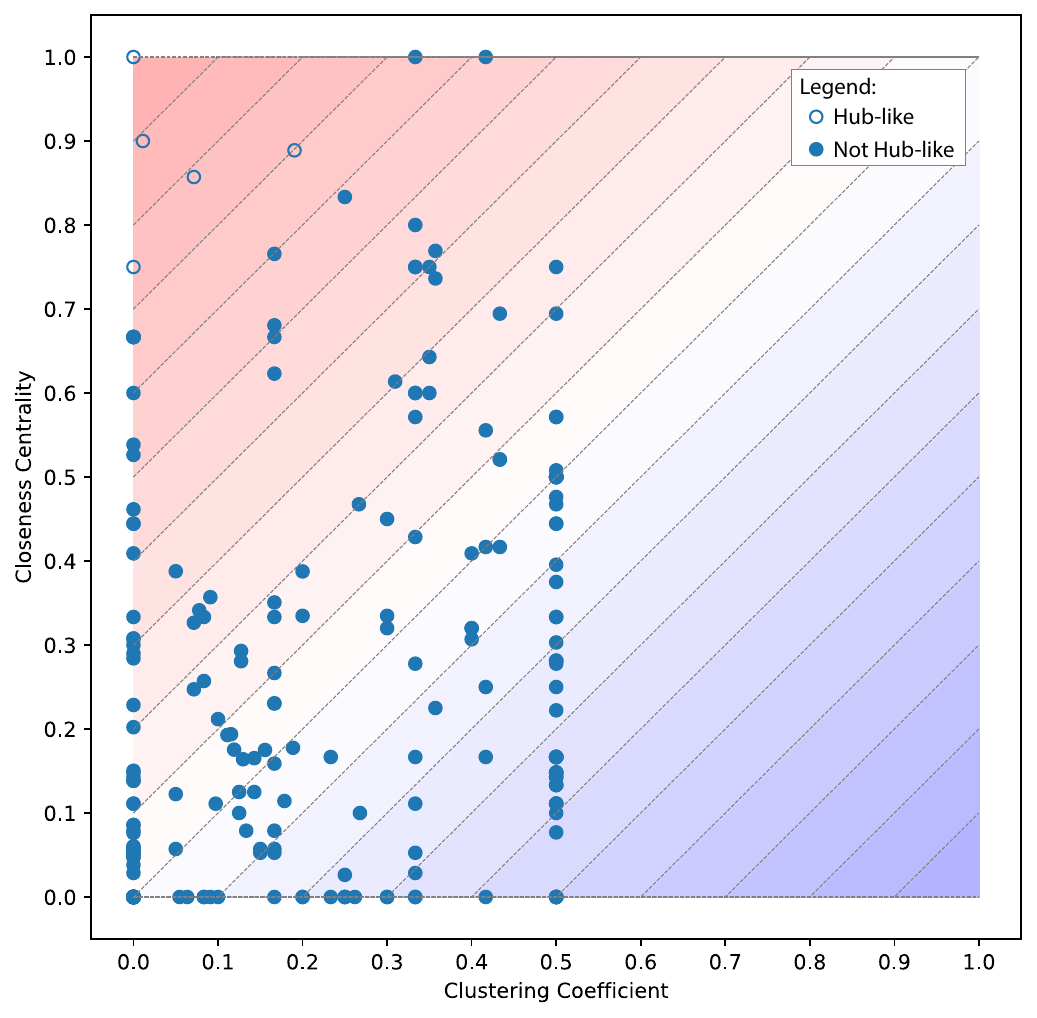}
    \caption{Scatter plot of Closeness Centrality and Clustering Coefficient. Diagonal lines represent areas where nodes have the same Hub strength.}
    \label{fig:closeness}
\end{figure}

Two main authors of this work examined the scatter plots of centrality and clustering coefficient for each considered centrality to \textbf{select Hub-like microservices} (e.g., Figure \ref{fig:closeness}).

We reached Excellent agreement (Table \ref{tab:kappa-agreement}) for Authority, Betweenness, Eigenvector, Out-degree, and PageRank; Good agreement for Closeness and degree centrality; Discrete agreement for in-degree centrality and Poor agreement for Hub centrality. The average Cohen's Kappa coefficient is $0.73$. We thus obtained \textbf{seven different lists of hub candidates} (Table \ref{tab:groups}).

\subsection{Arcan Approach (RQ$_{1.4}$)}

We implemented the threshold detection approach of Fontana et al. \cite{fontana2015automatic} used in the Arcan tool \cite{fontana2017arcan} and applied it to both the degree and degree centrality of the microservices. The obtained thresholds corresponding to the $0.75$ quartiles of the cropped distributions are 11 out of 12 and 0.67 out of 1.0, respectively. We considered the microservices to be Hub-like if they fall above the respective thresholds, and we call the corresponding approaches Arcan\_abs and Arcan\_norm. We \textbf{detected 10 hubs} according to Arcan\_abs and \textbf{66 hubs} according to Arcan\_norm (Table \ref{tab:groups}).

We can summarize the different sub-questions as:
\textbf{\emph{All considered methods provide candidates for Hub-like anti-pattern, while the studied networks are not scale-free.}
}

\subsection{Comparison of all methods (RQ$_{2}$)}
\begin{figure}
    \centering
    \includegraphics[width=\linewidth]{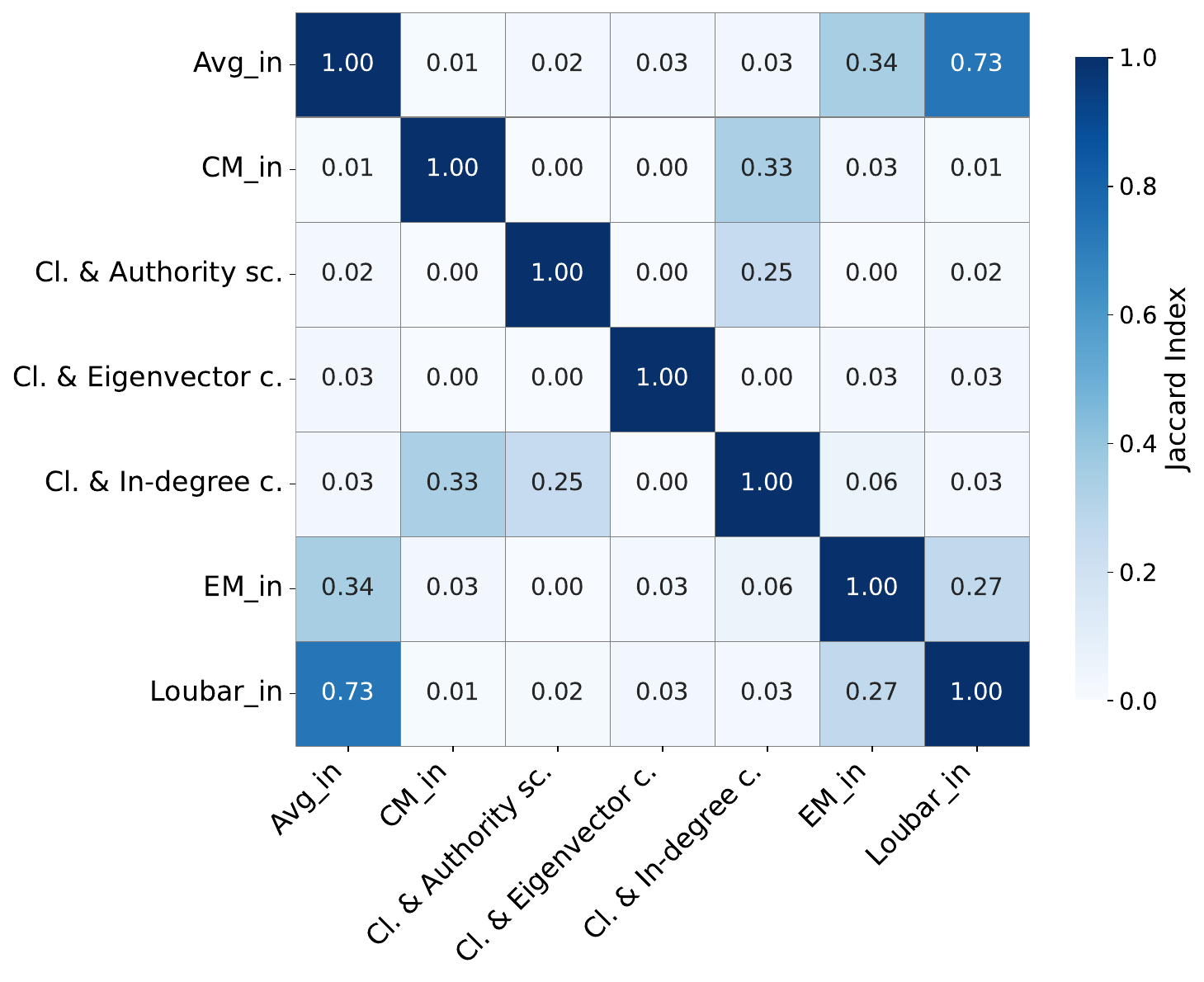}
    \caption{Jaccard index matrix for methods considering incoming connections only (Fleiss' Kappa 0.1003)}
    \label{fig:in}
\end{figure}

\begin{figure}
    \centering
    \includegraphics[width=\linewidth]{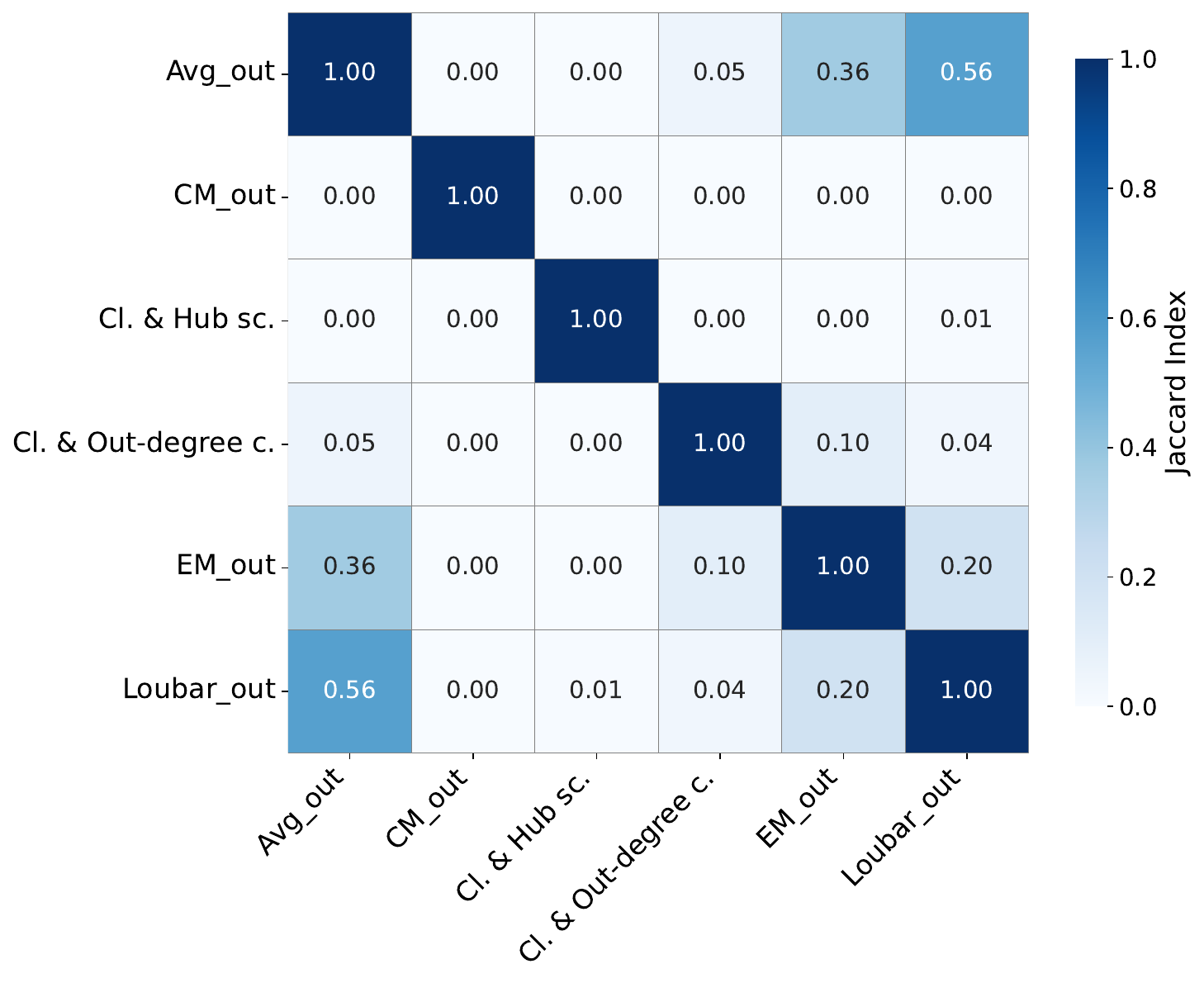}
    \caption{Jaccard index matrix for methods considering outgoing connections only (Fleiss' Kappa 0.065)}
    \label{fig:out}
\end{figure}

\begin{figure}
    \centering
    \includegraphics[width=\linewidth]{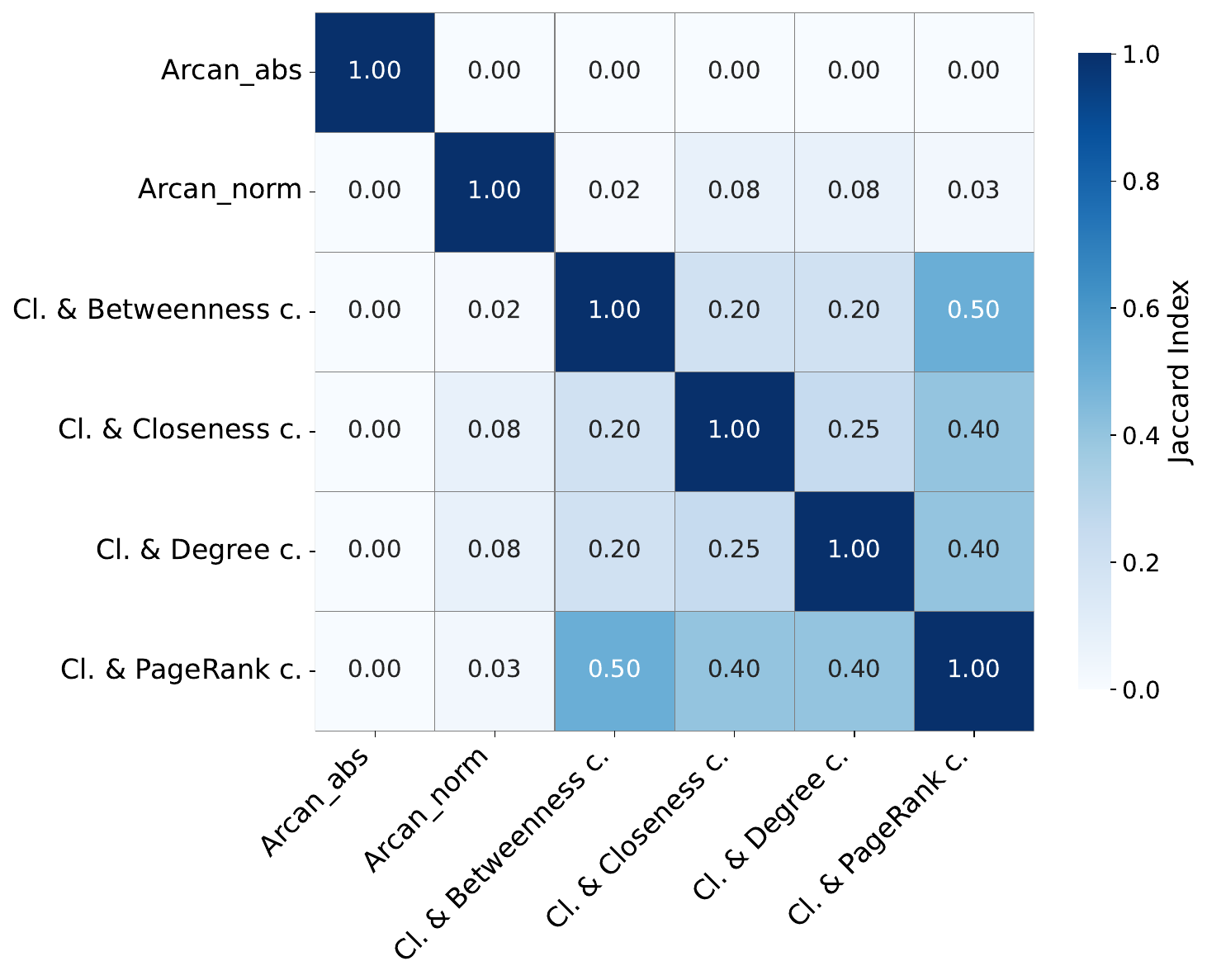}
    \caption{Jaccard index matrix for methods considering both the incoming and outgoing connections (Fleiss' Kappa 0.0423)}
    \label{fig:all}
\end{figure}

We have grouped the methods' results according to Table \ref{tab:groups}. The Jaccard coefficient matrices for the three groups of methods are shown in Figures \ref{fig:in}, \ref{fig:out}, and \ref{fig:all}.

The Fleiss Kappa values are $0.1003$ for detectors based on incoming connections, $0.065$ for outgoing connections, and $0.0423$ for detectors considering all possible connections. Further, considering all detectors altogether, the Fleiss Kappa is $0.0323$.
Generally, we can observe that \textbf{most methods disagree} on which services are Hub-like. Considering the methods utilizing incoming connections (Figure \ref{fig:in}), we see that the method coupling Clustering coefficient and \textbf{In-degree centrality has some overlap with the Authority score and CM\_in method}. Among the methods used by Kirkley, \textbf{Loubar, Avg, and EM all overlap with each other} in both cases, considering either the incoming or outgoing connections. Otherwise, \textbf{no other methods overlap when considering outgoing connections} (Figure \ref{fig:out}).

When it comes to methods considering all connections, we see that methods coupling the Clustering coefficient with \textbf{Betweenness, Closeness, Degree, and PageRank centralities have a noticeable overlap} (Figure \ref{fig:all}). However, we note that these methods yielded a low number of hubs (Table \ref{tab:precision}). Notably, \textbf{Arcan\_abs and Arcan\_norm do not overlap}.

We can thus answer the RQ$_{2}$ as follows:
\textbf{\emph{All considered methods provide distinct candidates for the Hub-like anti-pattern.}}

\subsection{Manual Validation (RQ$_{3}$)}

Out of 251 analyzed microservices, 209 were detected as a hub by at least one method (Table \ref{tab:precision}). Two main authors of this work have examined these services in their respective SDGs and judged whether they represent a hub. We achieved Good to Excellent agreement on each of the systems (Table \ref{tab:kappa-agreement}), with the average agreement being Good. After solving the disagreements, we concluded that \textbf{44 services are hubs}, \textbf{51 services are \emph{infrastructural} hubs}, and the remaining \textbf{114 services are not hubs} and thus represent False Positives. The Precision of each method, calculated while considering IH to be TP, FP, or ignoring them, is shown in Table \ref{tab:precision}.

We can make the following observations:

Out of the two methods utilizing the minimum description length principle (RQ$_{1.2}$), \textbf{CM did not identify any hubs} based on outgoing connections and only one infrastructural hub based on incoming connections, while \textbf{ER achieved reasonably good precision }on incoming connections, even when IH are not considered hubs while having lower precision when considering outgoing connections.  Notably, these two methods perform better than the two methods Kirkley used as reference~\cite{kirkley2024hubs}, namely the Average and Loubar methods, which detected a significant amount of nodes as hubs (Table \ref{tab:precision}).

When it comes to centrality metrics coupled with clustering coefficient (RQ$_{1.3}$), methods leveraging \textbf{Hubs score and Betweenness centrality did not capture a single hub} since their precision is zero in every case while methods leveraging \textbf{Authority score, In-degree, Our-degree, Degree, Closeness, PageRank centrality only detected \emph{infrastructural hubs}} since their precision is only non-zero if such services were considered True Positives. The method combining clustering coefficient and \textbf{Eigenvector centrality shows robust precision} with respect to IH, but with a \textbf{low precision} (0.2).

Concerning the performance of the two approaches replicating Arcan (RQ$_{1.4}$), it seems like Arcan\_abs has reached perfect precision. However, as mentioned in Data Analysis of Section \ref{sec:method}, the threshold of degree determined by Arcan\_abs was 11 out of 12, and very few services in our data have their degree above 10. In practice, only services from the two studied versions of \emph{train-ticket} system were detected by Arcan\_abs, thus demonstrating a drawback we anticipated in Sections \ref{sec:RW} and \ref{sec:method}. Thus, we can suspect that its Recall would be far from perfect. Conversely,  Arcan\_norm leveraged degree centrality, i.e., the normalized degree, and although demonstrating lower precision, identified hubs from 22 out of 25 studied MSAs. However, as with many other discussed methods, it \textbf{mostly identified \emph{infrastructural} hubs}.

Overall, we can also observe that \textbf{no method is capable of avoiding \emph{infrastructural} hubs} since the precision of all methods is highest when IH is TP. Moreover, \textbf{ER\_in has the best performance} with a reasonable number of detected hubs.

We can summarize the answer to RQ$_3$:
\textbf{\emph{Most methods detect \emph{infrastructural} hubs; the best method according to precision and number of detections is ER\_in.}}

\input{Tables/precision}

%% file: Tables/precision.tex
\begin{table}[]
\centering
\caption{Precision of hub detection approaches (number of detected hubs)}
\label{tab:precision}
\begin{tabular}{l|l|l|l} \hline 
                               & \textbf{IH are TP} & \textbf{No IH} & \textbf{IH are FP} \\ \hline
 \multicolumn{4}{l}{\textbf{Incoming connections}}                                                                       \\ \hline
Avg\_in (96) & 0.604      & 0.465   & 0.344                 \\
Loubar\_in (119)                  & 0.538      & 0.382  & 0.286      \\
CM\_in (1)                      & 1                     &            -          & 0                     \\
ER\_in   (33)                    & 0.848      & 0.783    & 0.545      \\
Cl. \& In-degree Centrality (3)   & 0.667      & 0                  & 0                     \\
Cl. \& Eigenvector Centrality (5) & 0.2                     & 0.2                  & 0.2                     \\
Cl. \& Authority Score   (2)     & 0.5                     & 0                  & 0                     \\ \hline
 \multicolumn{4}{l}{\textbf{Outgoing connections}}                                                                      \\ \hline
Avg\_out       (92)              & 0.598      & 0.413   & 0.283      \\
Loubar\_out       (138)           & 0.471     & 0.263  & 0.188     \\
CM\_out    (0)                  &               -          &           -           &         -                \\
ER\_out    (39)                  & 0.641      & 0.462  & 0.308      \\
Cl. \& Out-degree Centrality (5)  & 0.8                     & 0                  & 0                     \\
Cl. \& Hub Score         (1)     & 0                     & 0                  & 0                     \\ \hline
 \multicolumn{4}{l}{\textbf{All connections}}                                                                              \\ \hline
Arcan\_abs     (10)                      & 1                     & 1                  & 0.6                     \\
Arcan\_norm       (66)                   & 0.727      & 0.333   & 0.136     \\
Cl. \& Degree Centrality (5)      & 0.4                     & 0                  & 0                     \\
Cl. \& Betweenness Centrality (1) & 0                     & 0                  & 0                     \\
Cl. \& Closeness Centrality (5)  & 0.8                     & 0                  & 0                     \\
Cl. \& PageRank Centrality  (2)  & 0.5                     & 0                  & 0   \\ \hline                  
\end{tabular}
\end{table}

%% file: Sections/Discussion.tex

Firstly, the 25 studied networks were \textbf{not scale-free}, i.e., their degree distribution did not follow a power law. However, given the observed p-values and Figure \ref{fig:scalefree}, we could say that the networks are \textbf{\emph{almost} scale-free}, and the failed distribution test could be explained by a \textbf{small amount of data} and the scale of the networks available. We believe that a repeated investigation of the scale-free property of SDG of large industrial MSA projects should be carried out. A positive result would allow existing results and knowledge regarding scale-free networks from network science literature to be applied, like the existence of several hubs aggregating most of the nodes together \cite{barabasi2003scale, barabasi2009scale}.

Secondly, during the manual validation, we had to label almost all microservices as either hubs or infrastructural hubs in some networks. This goes against the common sense that only a few nodes in a network should be considered hubs, which aggregate mostly disconnected nodes, and affects the results of the validation. This issue is caused by the necessity to rely on \textbf{OSS microservice systems with few microservices}, since no large dataset of an MSA of an industrial system is currently available. Moreover, we identified more infrastructural hubs than proper hubs, which is another limitation of the used data. In practice, such services as \textbf{\textit{eureka, consul, kafka} should either be removed from the network before analysis or whitelisted from the Hub-like anti-pattern category}, since they are infrastructural tools required to connect to every other microservice by design and do not represent an architectural flaw. In this work, they had to be included both in the analyzed SDGs and the comparison and validation, because without them, the reconstructed networks would become disconnected, and we would have way fewer data points to work with. This could also be \textbf{resolved by studying an MSA of an industrial system}, since the amount of infrastructure service is limited, while the amount of microservices implementing business logic would be substantially higher.

Moreover, considering the outputs of each method, some of them provided a lot of hubs while some - very few. For instance, Loubar\_in provided 119 hubs out of 251 microservices (47\%), while Loubar\_out provided 138 (55\%) and Arcan\_norm - 66 (26\%). Such a \textbf{number of potential Hub-like microservices would be unfeasible} for an architect to evaluate, so the leveraged detection method should provide a reasonable, low number of detections. On the other hand, CM\_in and CM\_out failed to provide reasonable candidates, with CM\_out not providing any hubs and CM\_in detecting only an IH. Methods based on clustering coefficient and centrality metrics provide few hubs by design, since we focus on the cluster in the top-left corner of the scatter plot (Figure \ref{fig:closeness}). However, it is unknown what a plot would look like for an industrial MSA, and potentially, such a cluster would be larger. Overall, the \textbf{methods based on Erdős–Rényi encoding (ER\_in, ER\_out) seemed to provide a reasonable number of hubs}. Since the ER\_out performed worse than ER\_in, we can assume that the majority of Hubs in the studied SDGs are ones that make calls to the other services, and not the ones that are called by many services.

Furthermore, considering the agreement between the methods, we observe several trends in agreement and disagreement. Average and Loubar methods leveraged by Kirkley as benchmarks \cite{kirkley2024hubs} agree substantially (Figures \ref{fig:in} and \ref{fig:out}), which was expected since both methods apply different thresholds on the same metric - the degree of the node. \textbf{Methods leveraging the centrality and clustering coefficient do not agree with methods based on other approaches}. This further highlights the fact that \textbf{each centrality metric possesses different properties and prioritizes different nodes} (Section \ref{sec:back}), so the appropriate centrality metric needs to be selected for different analyses of the MSA \cite{abufouda2017using}. Among the centrality-based approaches, there seems to be some overlap between methods utilizing Closeness, Betweenness, PageRank, and degree centralities. This can be explained by the small scale of the networks - there are very few long paths in the studied networks, with most microservices calling each other directly or through chains of 2-3 services. This implies that the number of paths passing through a given node is comparable to the degree of the node, which explains the similarity in the detected hubs.

Interestingly, Arcan\_abs and Arcan\_norm have diverging results - \textbf{Arcan\_abs only detected hubs in the \emph{train-ticket}} system, while \textbf{Arcan\_norm detected hubs in most of the studied systems}, but had no overlap with Arcan\_abs. This result further corroborates the drawbacks and criticisms we outlined in Sections \ref{sec:RW} and \ref{sec:method}: using \textbf{absolute scales of the metrics can distort the results} if systems of different scales are analyzed, which was the case for our data. Moreover, the approach used by Arcan to detect the threshold is based on the previous research by the same authors \cite{fontana2015automatic}, and it contains several assumptions about the distribution of the studied metric, namely that the distribution has a long tail of low values and only the highest value have a significant peak. These assumptions serve as a basis for cutting the low-valued tail of the distribution and considering the quartiles of the remaining data. According to Arcan documentation, this approach is applied to the degree of a component to label it as Hub-like\footnote{\url{https://docs.arcan.tech/latest/architectural_smells/#hub-like-dependency}}. However, the degree of nodes in a network is unlikely to possess such properties \cite{kirkley2024hubs, barabasi2003scale}.

Finally, the \textbf{ER methods based on the Erdős–Rényi model} proposed by Kirkley \cite{kirkley2024hubs} appear to\textbf{ have a reasonable number of outputs, demonstrate agreement} with other established hub detection methods, and \textbf{achieved the best precision} through manual validation, when considering the performance with respect to infrastructural hubs.

The implication of our results for \emph{practitioners} is thus that there is reasonable evidence to consider utilizing either the normalized amount of connections of the microservice, i.e., the degree centrality, as the guiding metric for
detecting the Hub-like anti-pattern, or leveraging the ER hub detection method by Kirkley \cite{kirkley2024hubs} in the Arcan or similar tool, since these approaches enable more robust and generalizable analysis.

For \emph{researchers}, our work introduces several hub detection methods from the network science literature.
Investigating further the applicability of these methods to detecting Hub-like components in MSA and other systems opens up new research directions. 
In particular, we suggest further investigation of the applicability of the ER method for the assessment of microservice systems and the detection of Hub-like microservices. The Erdos–Rényi model was developed to model random networks, so its best performance for detecting Hub-like microservices suggests further investigation of whether microservice architectural networks compare to random networks.
Moreover, analyzing whether MSA networks possess the scale-free property requires further investigation on a large dataset.

%% file: Sections/Threats.tex
In this section, we discuss the threats to validity of this work according to guidelines by Wohlin \cite{wohlin_experimentation_2024}.

\textbf{Construct validity}.
Our data collection strategy can bias or skew the results. In this work, we leveraged the dataset provided by Bakhtin et al. \cite{bakhtin2025network}. The systems from this dataset are rather small, with the biggest one being \textit{train-ticket} with 42 microservices. This affects the performance of hub detection methods and manual validation. In particular, many nodes in the systems were considered hubs, and most of them are the infrastructural hubs. The only way to address and alleviate this threat is to repeat the study on a large industrial system.

\textbf{Internal validity}.
The goal of this work is to evaluate methods for hub detection in microservice networks. We selected applicable methods and used the existing implementations to perform data collection and analysis. Some of the methods considered, such as scale-free tests or betweenness and closeness centralities, are statistical in nature and are thus applicable to large networks, while we could only study rather small networks. To properly assess these methods, this study needs to be replicated on a large system.
When it comes to manual labeling, we employed standard procedures, such as measuring the agreement with the Cohen's Kappa coefficient.

\textbf{External validity}.
Due to the reliance on the dataset provided by Bakhtin et al. \cite{bakhtin2025network}, our analysis and conclusions are based on OSS microservice benchmarks and not on the industrial systems. We were thus careful not to overgeneralize our findings and focused on the performance of the hub detection methods on the studied data. We highlighted several promising methods that practitioners and researchers should consider further.

\textbf{Conclusion validity}.
We base our conclusions on the manual validation of Hub-like microservices performed by two authors. Due to concerns of construct validity, the networks contained many hubs, with a significant portion of them being infrastructural hubs. Moreover, the definition of the Hub-like anti-pattern is ambiguous and subjective, and we do not know how practitioners perceive this anti-pattern. Thus, the results of the validation might be biased towards the researchers' point of view.

%% file: Sections/Conclusion.tex
In this work, we evaluated several methods for detecting the Hub-like microservice anti-pattern based on network science concepts on a dataset of 25 SDGs.
Our findings suggest the adoption of degree centrality instead of the absolute degree of a microservice in the SDG to reason about the topology of the network and detect the Hub-like anti-pattern. Moreover, the method by Kirkley \cite{kirkley2024hubs} based on the Erdős–Rényi model proved to be the most robust and precise.

Data and scripts to replicate the study are available in the Replication Package \cite{replication}.

%% file: main.bbl
\begin{thebibliography}{10}
\providecommand{\url}[1]{#1}
\csname url@samestyle\endcsname
\providecommand{\newblock}{\relax}
\providecommand{\bibinfo}[2]{#2}
\providecommand{\BIBentrySTDinterwordspacing}{\spaceskip=0pt\relax}
\providecommand{\BIBentryALTinterwordstretchfactor}{4}
\providecommand{\BIBentryALTinterwordspacing}{\spaceskip=\fontdimen2\font plus
\BIBentryALTinterwordstretchfactor\fontdimen3\font minus \fontdimen4\font\relax}
\providecommand{\BIBforeignlanguage}[2]{{%
\expandafter\ifx\csname l@#1\endcsname\relax
\typeout{** WARNING: IEEEtran.bst: No hyphenation pattern has been}%
\typeout{** loaded for the language `#1'. Using the pattern for}%
\typeout{** the default language instead.}%
\else
\language=\csname l@#1\endcsname
\fi
#2}}
\providecommand{\BIBdecl}{\relax}
\BIBdecl

\bibitem{lewis14_microservices}
\BIBentryALTinterwordspacing
J.~Lewis and M.~Fowler, ``Microservices: a definition of this new architectural term,'' 2014. [Online]. Available: \url{https://martinfowler.com/articles/microservices.html}
\BIBentrySTDinterwordspacing

\bibitem{bakhtin2022survey}
A.~Bakhtin, A.~Al~Maruf, T.~Cerny, and D.~Taibi, ``Survey on tools and techniques detecting microservice api patterns,'' in \emph{2022 IEEE International Conference on Services Computing (SCC)}.\hskip 1em plus 0.5em minus 0.4em\relax IEEE, 2022, pp. 31--38.

\bibitem{cerny2023catalog}
T.~Cerny, A.~S. Abdelfattah, A.~Al~Maruf, A.~Janes, and D.~Taibi, ``Catalog and detection techniques of microservice anti-patterns and bad smells: A tertiary study,'' \emph{Journal of Systems and Software}, vol. 206, p. 111829, 2023.

\bibitem{pigazzini2020towards}
I.~Pigazzini, F.~A. Fontana, V.~Lenarduzzi, and D.~Taibi, ``Towards microservice smells detection,'' in \emph{Proceedings of the 3rd International Conference on Technical Debt}, 2020, pp. 92--97.

\bibitem{al2022using}
A.~Al~Maruf, A.~Bakhtin, T.~Cerny, and D.~Taibi, ``Using microservice telemetry data for system dynamic analysis,'' in \emph{2022 IEEE International Conference on Service-Oriented System Engineering (SOSE)}.\hskip 1em plus 0.5em minus 0.4em\relax IEEE, 2022, pp. 29--38.

\bibitem{bakhtin2022microservice}
A.~Bakhtin, ``Microservice api pattern detection : Using business processes and call graphs,'' Master's thesis, Tampere University, 2022.

\bibitem{bakhtin2025network}
A.~Bakhtin, M.~Esposito, V.~Lenarduzzi, and D.~Taibi, ``Network centrality as a new perspective on microservice architecture,'' in \emph{2025 IEEE 22nd International Conference on Software Architecture (ICSA)}, 2025, pp. 72--83.

\bibitem{kirkley2024hubs}
A.~Kirkley, ``Identifying hubs in directed networks,'' \emph{Phys. Rev. E}, vol. 109, p. 034310, Mar 2024.

\bibitem{azadi2019architectural}
U.~Azadi, F.~A. Fontana, and D.~Taibi, ``Architectural smells detected by tools: a catalogue proposal,'' in \emph{2019 IEEE/ACM International Conference on Technical Debt (TechDebt)}.\hskip 1em plus 0.5em minus 0.4em\relax IEEE, 2019, pp. 88--97.

\bibitem{barabasi2003scale}
A.-L. Barab{\'a}si and E.~Bonabeau, ``Scale-free networks,'' \emph{Scientific american}, vol. 288, no.~5, pp. 50--9, 2003.

\bibitem{barabasi2009scale}
A.-L. Barab{\'a}si, ``Scale-free networks: a decade and beyond,'' \emph{science}, vol. 325, no. 5939, pp. 412--413, 2009.

\bibitem{fontana2015automatic}
F.~A. Fontana, V.~Ferme, M.~Zanoni, and A.~Yamashita, ``Automatic metric thresholds derivation for code smell detection,'' in \emph{2015 IEEE/ACM 6th International Workshop on Emerging Trends in Software Metrics}.\hskip 1em plus 0.5em minus 0.4em\relax IEEE, 2015, pp. 44--53.

\bibitem{fontana2017arcan}
F.~A. Fontana, I.~Pigazzini, R.~Roveda, D.~Tamburri, M.~Zanoni, and E.~Di~Nitto, ``Arcan: A tool for architectural smells detection,'' in \emph{2017 IEEE International Conference on Software Architecture Workshops (ICSAW)}.\hskip 1em plus 0.5em minus 0.4em\relax IEEE, 2017, pp. 282--285.

\bibitem{palma2015study}
F.~Palma and N.~Mohay, ``A study on the taxonomy of service antipatterns,'' in \emph{2015 IEEE 2nd International Workshop on Patterns Promotion and Anti-patterns Prevention (PPAP)}.\hskip 1em plus 0.5em minus 0.4em\relax IEEE, 2015, pp. 5--8.

\bibitem{walker2020automated}
A.~Walker, D.~Das, and T.~Cerny, ``Automated code-smell detection in microservices through static analysis: A case study,'' \emph{Applied Sciences}, vol.~10, no.~21, p. 7800, 2020.

\bibitem{walker2021automated}
------, ``Automated microservice code-smell detection,'' in \emph{Information Science and Applications: Proceedings of ICISA 2020}.\hskip 1em plus 0.5em minus 0.4em\relax Springer, 2021, pp. 211--221.

\bibitem{bushong2021microservice}
V.~Bushong, A.~S. Abdelfattah, A.~A. Maruf, D.~Das, A.~Lehman, E.~Jaroszewski, M.~Coffey, T.~Cerny, K.~Frajtak, P.~Tisnovsky \emph{et~al.}, ``On microservice analysis and architecture evolution: A systematic mapping study,'' \emph{Applied Sciences}, vol.~11, no.~17, p. 7856, 2021.

\bibitem{das2022technical}
D.~Das, A.~A. Maruf, R.~Islam, N.~Lambaria, S.~Kim, A.~S. Abdelfattah, T.~Cerny, K.~Frajtak, M.~Bures, and P.~Tisnovsky, ``Technical debt resulting from architectural degradation and code smells: a systematic mapping study,'' \emph{ACM SIGAPP Applied Computing Review}, vol.~21, no.~4, pp. 20--36, 2022.

\bibitem{cerny2022microvision}
T.~Cerny, A.~S. Abdelfattah, V.~Bushong, A.~Al~Maruf, and D.~Taibi, ``Microvision: Static analysis-based approach to visualizing microservices in augmented reality,'' in \emph{2022 IEEE International Conference on Service-Oriented System Engineering (SOSE)}.\hskip 1em plus 0.5em minus 0.4em\relax IEEE, 2022, pp. 49--58.

\bibitem{huizinga2023detecting}
A.~Huizinga, G.~Parker, A.~S. Abdelfattah, X.~Li, T.~Cerny, and D.~Taibi, ``Detecting microservice anti-patterns using interactive service call graphs: Effort assessment,'' in \emph{Southwest Data Science Conference}.\hskip 1em plus 0.5em minus 0.4em\relax Springer, 2023, pp. 212--227.

\bibitem{gortney2022visualizing}
M.~E. Gortney, P.~E. Harris, T.~Cerny, A.~Al~Maruf, M.~Bures, D.~Taibi, and P.~Tisnovsky, ``Visualizing microservice architecture in the dynamic perspective: A systematic mapping study,'' \emph{IEEE Access}, vol.~10, pp. 119\,999--120\,012, 2022.

\bibitem{cerny2024static}
T.~Cerny, A.~S. Abdelfattah, J.~Yero, and D.~Taibi, ``From static code analysis to visual models of microservice architecture,'' \emph{Cluster Computing}, vol.~27, no.~4, pp. 4145--4170, 2024.

\bibitem{sharma2016designite}
T.~Sharma, P.~Mishra, and R.~Tiwari, ``Designite: A software design quality assessment tool,'' in \emph{Proceedings of the 1st international workshop on bringing architectural design thinking into developers' daily activities}, 2016, pp. 1--4.

\bibitem{le2016relating}
D.~M. Le, C.~Carrillo, R.~Capilla, and N.~Medvidovic, ``Relating architectural decay and sustainability of software systems,'' in \emph{2016 13th Working IEEE/IFIP Conference on Software Architecture (WICSA)}.\hskip 1em plus 0.5em minus 0.4em\relax IEEE, 2016, pp. 178--181.

\bibitem{capilla2023detecting}
R.~Capilla, F.~A. Fontana, T.~Mikkonen, P.~Bacchiega, and V.~Salamanca, ``Detecting architecture debt in micro-service open-source projects,'' in \emph{2023 49th Euromicro Conference on Software Engineering and Advanced Applications (SEAA)}, 2023, pp. 394--401.

\bibitem{bacchiega2024refactoring}
P.~Bacchiega, D.~Rusconi, P.~Mereghetti, and F.~A. Fontana, ``Refactoring of a microservices project driven by architectural smell detection,'' in \emph{2024 IEEE 21st International Conference on Software Architecture Companion (ICSA-C)}.\hskip 1em plus 0.5em minus 0.4em\relax IEEE, 2024, pp. 281--288.

\bibitem{pigazzini2022exploiting}
I.~Pigazzini, D.~Di~Nucci, F.~A. Fontana, and M.~Belotti, ``Exploiting dynamic analysis for architectural smell detection: a preliminary study,'' in \emph{2022 48th Euromicro Conference on Software Engineering and Advanced Applications (SEAA)}.\hskip 1em plus 0.5em minus 0.4em\relax IEEE, 2022, pp. 282--289.

\bibitem{lino2024musvision}
J.~F. Lino~Daniel, M.~Vidu, T.~Rosa, A.~Goldman, and E.~Guerra, ``Msvision: An extensible tool for detecting patterns and bad smells in heterogeneous microservices systems,'' \emph{Available at SSRN 4892088}, 2024.

\bibitem{silva2024mfe}
\BIBentryALTinterwordspacing
N.~Silva, E.~Rodrigues, and T.~Conte, ``A catalog of micro frontends anti-patterns,'' 2024. [Online]. Available: \url{https://arxiv.org/abs/2411.19472}
\BIBentrySTDinterwordspacing

\bibitem{watt2019using}
N.~Watt, ``{Using Graph Theory and Network Science to Explore Your Microservices Architecture},'' in \emph{Proceedings of GOTO Berlin 2019}, 2019.

\bibitem{engel2018evaluation}
T.~Engel, M.~Langermeier, B.~Bauer, and A.~Hofmann, ``Evaluation of microservice architectures: A metric and tool-based approach,'' in \emph{Information Systems in the Big Data Era: CAiSE Forum 2018, Tallinn, Estonia, June 11-15, 2018, Proceedings 30}.\hskip 1em plus 0.5em minus 0.4em\relax Springer, 2018, pp. 74--89.

\bibitem{gaidels2020service}
E.~Gaidels and M.~Kirikova, ``Service dependency graph analysis in microservice architecture,'' in \emph{Perspectives in Business Informatics Research: 19th International Conference on Business Informatics Research, BIR 2020, Vienna, Austria, September 21--23, 2020, Proceedings 19}.\hskip 1em plus 0.5em minus 0.4em\relax Springer International Publishing, 2020, pp. 128--139.

\bibitem{farsi:hal-03825330}
\BIBentryALTinterwordspacing
H.~Farsi, ``{Industry 4.0 and Microservices with the power of Graphs},'' {International Workshop of Services and Industry of the Future (IWSIF 2022)}, Oct. 2022, poster. [Online]. Available: \url{https://hal.science/hal-03825330}
\BIBentrySTDinterwordspacing

\bibitem{borges2019algorithm}
R.~Borges and T.~Khan, ``Algorithm for detecting antipatterns in microservices projects,'' \emph{CEUR workshop proceedings}, 2019.

\bibitem{rud2006product}
D.~Rud, A.~Schmietendorf, and R.~R. Dumke, ``Product metrics for service-oriented infrastructures,'' \emph{IWSM/MetriKon}, pp. 161--174, 2006.

\bibitem{albert1999diameter}
R.~Albert, H.~Jeong, and A.-L. Barab{\'a}si, ``Diameter of the world-wide web,'' \emph{nature}, vol. 401, no. 6749, pp. 130--131, 1999.

\bibitem{lavazza2016identifying}
L.~Lavazza and S.~Morasca, ``Identifying thresholds for software faultiness via optimistic and pessimistic estimations,'' in \emph{Proceedings of the 10th ACM/IEEE International Symposium on Empirical Software Engineering and Measurement}, 2016, pp. 1--10.

\bibitem{morasca2016slope}
S.~Morasca and L.~Lavazza, ``Slope-based fault-proneness thresholds for software engineering measures,'' in \emph{Proceedings of the 20th international conference on evaluation and assessment in software engineering}, 2016, pp. 1--10.

\bibitem{bakhtin2025ccp}
\BIBentryALTinterwordspacing
A.~Bakhtin, M.~Esposito, V.~Lenarduzzi, and D.~Taibi, ``Centrality change proneness: an early indicator of microservice architectural degradation,'' 2025. [Online]. Available: \url{https://arxiv.org/abs/2506.07690}
\BIBentrySTDinterwordspacing

\bibitem{esposito2023can}
M.~Esposito, S.~Moreschini, V.~Lenarduzzi, D.~H{\"a}stbacka, and D.~Falessi, ``Can we trust the default vulnerabilities severity?'' in \emph{2023 IEEE 23rd International Working Conference on Source Code Analysis and Manipulation (SCAM)}.\hskip 1em plus 0.5em minus 0.4em\relax IEEE, 2023, pp. 265--270.

\bibitem{falessi2023enhancing}
D.~Falessi, S.~M. Laureani, J.~{\c{C}}arka, M.~Esposito, and D.~A.~d. Costa, ``Enhancing the defectiveness prediction of methods and classes via jit,'' \emph{Empirical Software Engineering}, vol.~28, no.~2, p.~37, 2023.

\bibitem{wohlin_experimentation_2024}
C.~Wohlin, P.~Runeson, M.~H{\"{o}}st \emph{et~al.}, \emph{Experimentation in Software Engineering}.\hskip 1em plus 0.5em minus 0.4em\relax Springer, 2012.

\bibitem{newman2001clustering}
M.~E. Newman, ``Clustering and preferential attachment in growing networks,'' \emph{Physical review E}, vol.~64, no.~2, p. 025102, 2001.

\bibitem{Code2DFD23}
S.~Schneider and R.~Scandariato, ``Automatic extraction of security-rich dataflow diagrams for microservice applications written in java,'' \emph{Journal of Systems and Software}, vol. 202, p. 111722, 2023.

\bibitem{schneider2024comparison}
S.~Schneider, A.~Bakhtin, X.~Li, J.~Soldani, A.~Brogi, T.~Cerny, R.~Scandariato, and D.~Taibi, ``Comparison of static analysis architecture recovery tools for microservice applications,'' \emph{Empirical Software Engineering}, 2025.

\bibitem{imranur2019curated}
M.~I. Rahman, S.~Panichella, and D.~Taibi, ``A curated dataset of microservices-based systems,'' \emph{arXiv preprint arXiv:1909.03249}, 2019.

\bibitem{schneider2023microsecend}
S.~Schneider, T.~{\"O}zen, M.~Chen, and R.~Scandariato, ``microsecend: A dataset of security-enriched dataflow diagrams for microservice applications,'' in \emph{2023 IEEE/ACM 20th International Conference on Mining Software Repositories (MSR)}.\hskip 1em plus 0.5em minus 0.4em\relax IEEE, 2023, pp. 125--129.

\bibitem{amoroso2024dataset}
D.~A. d'Aragona, A.~Bakhtin, X.~Li, R.~Su, L.~Adams, E.~Aponte, F.~Boyle, P.~Boyle, R.~Koerner, J.~Lee \emph{et~al.}, ``A dataset of microservices-based open-source projects,'' in \emph{Proceedings of the 21st International Conference on Mining Software Repositories}, 2024, pp. 504--509.

\bibitem{yang2024feature}
W.~Yang, B.~Song, Y.~Xing, Y.~Lyu, H.~Cui, Z.~Liang, and Z.~Tu, ``A feature dataset of microservices-based systems,'' in \emph{International Conference on Service Science}.\hskip 1em plus 0.5em minus 0.4em\relax Springer, 2024, pp. 73--87.

\bibitem{replication}
\BIBentryALTinterwordspacing
A.~Bakhtin, M.~Esposito, V.~Lenarduzzi, and D.~Taibi, ``Replication package for “leveraging network methods for hub-like microservice detection”,'' May 2025. [Online]. Available: \url{https://zenodo.org/doi/10.5281/zenodo.15362705}
\BIBentrySTDinterwordspacing

\bibitem{zhou2005maximal}
T.~Zhou, G.~Yan, and B.-H. Wang, ``Maximal planar networks with large clustering coefficient and power-law degree distribution,'' \emph{Physical Review E—Statistical, Nonlinear, and Soft Matter Physics}, vol.~71, no.~4, p. 046141, 2005.

\bibitem{cohen_coefficient_1960}
J.~Cohen, ``A coefficient of agreement for nominal scales,'' \emph{Educational and Psychological Measurement}, vol.~20, no.~1, pp. 37--46, 1960, publisher: SAGE Publications.

\bibitem{falotico2015fleiss}
R.~Falotico and P.~Quatto, ``Fleiss’ kappa statistic without paradoxes,'' \emph{Quality \& Quantity}, vol.~49, pp. 463--470, 2015.

\bibitem{abufouda2017using}
M.~Abufouda and H.~Abukwaik, ``On using network science in mining developers collaboration in software engineering: A systematic literature review,'' \emph{International Journal of Data Mining \& Knowledge Management Process (IJDKP)}, vol.~7, no. 5/6, pp. 17--34, November 2017.

\end{thebibliography}
